\documentclass[12pt]{iopart}
\usepackage{color}
\usepackage{iopams}
\usepackage{amsfonts}
\usepackage{graphicx} 
\usepackage{amsopn}
\usepackage{citesort}
\DeclareMathOperator{\sech}{{sech }}
\DeclareMathOperator{\csch}{{csch }}
\newcommand{\dps}{{\displaystyle}}

\newcommand{\cL}{{\cal L}}
\newcommand{\cK}{{\cal K}}
\newcommand{\BBR}{{\mathbb{R}}}
\begin{document}
\title{Multiple-scale analysis on the radiation within the nonlinearly coupled KdV equations}

\author{Yezheng Li$^1$
, Joseph A. Biello$^2$, Yerong Li$^3$}
\address{$^1$ Applied Mathematics and Computational Science, University of Pennsylvania, David Rittenhouse Laboratory, 209 South 33rd Street, Philadelphia, PA 19104.}
\address{$^2$ Department of Mathematics, University of California, Davis, CA 95616.}
\address{$^3$ Department of Applied Physics and Materials Science, California Institute of Technology, Pasadena, CA 91125, USA}
\eads{\mailto{yezheng@sas.upenn.edu}, \mailto{biello@math.ucdavis.edu}, \mailto{yerong.li@caltech.edu}}
\begin{abstract}
A multiple scale model of the nonlinearly coupled KdV equations is established to predict mechanism of interaction of equatorial Rossby waves and barotropic waves in certain case. Analytically, predicted precursor radiation is  a centrosymmetric object and is shown in excellent quantitative agreement with numerical simulations; furthermore, the multiple scale model elucidates the salient mechanisms of the interaction of solitary waves and the mechanism for radiation. While the atmosphere-ocean science community is very interested in theoretical studies of tropical wave interactions and in developing reduced dynamical models that can explain some key features of equatorial phenomena, our analytic predictions quantitively explain formation of radiation during interaction in Biello's model beyond qualitative level.
\end{abstract}


\submitto{\NL}

\maketitle

\section{Introduction}
\subsection{Atmospheric background and solitary structures}
Some poorly understood phenomena in atmosphere-ocean science involve a complex nonlinear interaction of clouds, moisture, and convection on a large variety of scales in both time and space, ranging from cumulus clouds over a few kilometers to intraseasonal oscillations over planetary scales of order 40,000 km \cite{biello2005new,frierson2004large,majda2004multiscale,smith2013physics}. While current numerical simulations still fail to capture mechanism of interaction on multiple scales \cite{biello2005new,majda2004multiscale,majda2003systematic}, ocean-atmosphere science community is interested in theoretical studies of tropical wave interactions and in developing reduced dynamical models that can  at least qualitatively explain some key features of equatorial phenomena \cite{biello2004effect,biello2005new,frierson2004large,majda2003nonlinear,majda2004multiscale,majda2003systematic,majda1999interaction,dutrifoy2009simple}.

Under this circumstance, amptitude equations describing the interaction of equatorial Rossby waves and barotropic waves
\begin{eqnarray}
a_t -(1-2\gamma) a_{xxx} +(ab)_x && = 0,\\
b_t - b_{xxx} + \left(\frac{a^2}2\right)_x && = 0,
\label{eq:biello's amptitude equations}
\end{eqnarray}
 are derived by Biello and Majda \cite{biello2004effect,majda2003nonlinear} and are used as a model for long range interactions (teleconnections) between the tropical and midlatitude troposphere. On one hand, assessing the accuracy of this low-dimensional model (amongst the many) of the tropical atmosphere that take advantage of equatorial long-wave theory \cite{majda2003nonlinear,majda2009skeleton,stechmann2014walker} is an important ongoing task \cite{stechmann2015identifying,ogrosky2015mjo,ogrosky2015assessing}; on the other hand, different from some test models without instability nor positive Lyapunov exponents \cite{harlim2013test}, this simplified quasi-equilibrium tropical model (\ref{eq:biello's amptitude equations}) is more realistic achieved by allowing active barotropic dynamics and coupled nonlinear advection which allows for tropical-extratropical wave interactions \cite{majda2003nonlinear,biello2004boundary,biello2004effect,harlim2013test,khouider2005non}. 

 \cite{biello2004effect,biello2009nonlinearly} also explore solitary structures (which are not rare in atmospheric science) within these amplitude equations (\ref{eq:biello's amptitude equations}) which may explain transfer of energy between waves. The nonlinearly coupled KdV equations
\begin{eqnarray*}
u_t-(1-\gamma)u_{xxx}+uu_x=\gamma v_{xxx}+\frac{\left(uv\right)_x}{2}, \nonumber \\
v_t-(1-\gamma)v_{xxx}+vv_x=\gamma u_{xxx}+\frac{\left(uv\right)_x}{2}.
\end{eqnarray*}
are actually recast by a linear transformation from these amplitude equations (\ref{eq:biello's amptitude equations}) \cite{biello2009nonlinearly}. $u,v$ govern the amplitude of two types of modes, each of which consists of a coupled tropical/midlatitude flow. We choose to work with $\gamma=0$
\begin{eqnarray}
u_t-u_{xxx}+uu_x=\frac{\left(uv\right)_x}{2}, \nonumber \\
v_t-v_{xxx}+vv_x=\frac{\left(uv\right)_x}{2}.
\label{eq:nonlinearly coupled KdV equation of Biello}
\end{eqnarray}
due to the generic nature of the interaction \cite{biello2009nonlinearly}. 

Despite the need to understand its role in atmospheric background \cite{ji2014interhemispheric,ji2015nino,harlim2013test,raupp2010interaction}, analysis of solitons' behavior is very incomplete and it is precisely at the point of solitary waves that interesting dynamics arise \cite{biello2004effect}. Notice $u\equiv 0$ or  $v\equiv 0$ are invariant subspaces of (\ref{eq:nonlinearly coupled KdV equation of Biello}) with the non-zeron function evolving according to the KdV equation \cite{biello2009nonlinearly}; take $v\equiv 0$ as an example, KdV solution includes 
\begin{displaymath}
u(x,t)=\epsilon^{-2}K\left(\epsilon^{-1}\left(x-c\epsilon^{-2}t\right)\right), \qquad K(\xi)=-12\sech^2\xi, \qquad c=-4.
\end{displaymath}
and $c\epsilon^{-2}$ is the traveling velocity of the soliton.
Inspired by these solitary structures, Biello \cite{biello2009nonlinearly} presents numerical results showing the collision of one $u$-soliton with one $v$-soliton initialized by 
\begin{equation}
u(x,0)= K(x), \qquad v(x,0)=\epsilon^{-2}K\left(\epsilon^{-1}\left(x-x^0_v\right)\right),
 \label{eq:initial data}
\end{equation}
but only with relatively large $\epsilon =0.75$. Indeed, two solitons remain their solitary form before they come close to each other and interaction happens later on even though such initialization is well-posed due to work \cite{li2015global,jerry2014global,guo2013global}. Just by numerical simulation majorly with $\epsilon=0.75$, Biello \cite{biello2009nonlinearly} is sharp enough to insinuate some details of the interaction including shears, a small amount of rightward traveling radiation generated, etc., and a broad range of interesting and unexplained of behavior is displayed in just few numerical examples \cite{biello2009nonlinearly}. As a typical phenomenon in the simulation, the transversely narrow and sharply peaked soliton shape results in an eye-catching but small scale centrosymmetric precursor radiation in $v$. Such a peak has been observed experimentally in figure \ref{fig:V2, order 1 of u,v} and provides a possible mechanism for the formation of shear and radiation show in figure \ref{fig:snapshots of U1 scrutine},\ref{fig:V2, order 1 of u,v}; \cite{biello2009nonlinearly} also presents analogous shear, peak, radiation. 
\subsection{Multiple-scale model and approximation results} 
\label{sec:{Multiple-scale model and approximation results} }

 In this paper, we focus on the case of $\epsilon \to 0$ in (\ref{eq:initial data}). Notably, numerical simulation of solitary interaction is both our start and our end for solitons' behavior studying. Speaking of simulations, radiations during and after interaction of solitons pose a particular difficulty for theoretical investigation and numerical simulations of solitary waves representing "soliton amplitudes". Eventually, we successfully apply multiple-scale model to analytically predict radiations, interaction (with $\epsilon \to 0$):
\begin{enumerate}
\item  the ansatz of multiple scale model is
\begin{eqnarray*}
u(x,\epsilon^{-1}x,\epsilon^{-3}t)= && U^0(x) +U^{F}(x,\epsilon^{-1}x,\epsilon^{-3}t)\nonumber\\
&& +\epsilon U^{B}(x,\epsilon^{-1}x,\epsilon^{-3}t)+o(\epsilon), \nonumber \\
v(x,\epsilon^{-1}x,\epsilon^{-3}t)= && \epsilon^{-2} V^0(\epsilon^{-1}x,\epsilon^{-3}t) +V^{F}(x,\epsilon^{-1}x,\epsilon^{-3}t)\nonumber\\
&& +\epsilon V^{R}(x,\epsilon^{-1}x,\epsilon^{-3}t)+o(\epsilon); 
\end{eqnarray*}
where we have two scales in space and a single scale in time; or alternatively
\begin{eqnarray}
u(x,X,T) && =U^0(x) +U^{F}(x,X,T)+\epsilon U^{B}(x,X,T)+o(\epsilon),\nonumber \\
v(x,X,T) && =\epsilon^{-2}V^0(X,T)+V^{F}(x,X,T)+\epsilon V^{R}(x,X,T) +o(\epsilon).
\label{eq:ansatz}
\end{eqnarray}
where small space scale is $X=\epsilon^{-1} x$ and the single time scale is $T=\epsilon^{-3} t$ (fast time).

$U^{0}$ is the base of wave $u$ and $V^{0}$ is approximately the original sharp soliton of wave $v$. 
$U^{F}$ (figure \ref{fig:observation of U1}) is fleeting solitary structure inside wave $u$ while (inconspicuous) $V^{F}$ is the counterpart for wave $v$. $\epsilon V^{R}$ is precursor $v$ radiation (figure \ref{fig:observation of V2 descry}, \ref{fig:V2, order 1 of u,v}), one of our focuses of attention while (less important and obscure as well) $\epsilon U^{B}$ is the bruise left on $u$ after the interaction.
\item the model captures mechanism (three phases of interaction) for the formation of shear and radiation: 
\begin{quote}
precursor $v$-radiation is formed firstly; then it interacts with $u$, forming $u$-radiation and $v$-radiation in-situ.
\end{quote}
\item the precursor $v$ radiation (after the interaction) is of the form 
\begin{equation}
\epsilon V^{R}(x,X,T) = \epsilon  C^{R}_VR(x), \mbox{ as } T\to \infty,
\label{eq:conclusion of precursor v radiation in the introduction}
\end{equation}
where normalized $R(x) = \frac{3\sqrt{3}}2\sinh(x)\cosh^3(x)$ and
$C^{R}_V$ is a positive constant representing height of $V^{R}$: $C^{R}_V = 1.03\sim 1.14$ from simulation (figure \ref{fig:observation of V2 descry},\ref{fig:V2, order 1 of u,v},\ref{fig:heaviside structure H2}; table \ref{tab:theoretical maximum V1 V2}) and our WKB model predicts $C^{R}_V =4.11$ (second part of figure \ref{fig:G2 H2}; table \ref{tab:theoretical maximum V1 V2}; derivation in (\ref{eq:right boundary value for H2 use integral of H})). By comparison, our WKB model's prediction of less important $U^{B}$ agrees extremely well with numerical simulations with $C^{B}_U = -15.37$ (figure \ref{fig:snapshots of U1 scrutine},\ref{fig:heaviside structure G2}, first part of figure \ref{fig:G2 H2}; table \ref{tab:theoretical maximum U1 U2}; derivation in (\ref{eq:right boundary value for G2 use integral of G})) within $\epsilon U^{B}(x,X,T) = \epsilon C^{B}_U R(x)$, as $T\to \infty$.
\end{enumerate}

\subsection{Outline of solution strategy}
\label{sec:outline of the paper}

The objective of this work is to use multiple scale model to predict "precursor" radiation $\epsilon V^{R}$ (figure \ref{fig:observation of V2 descry},\ref{fig:V2, order 1 of u,v}) as $\epsilon\to 0$ in (\ref{eq:nonlinearly coupled KdV equation of Biello},\ref{eq:initial data}). Our analytical strategy has a few key steps which exploit the multi-scale structure observed from numerical simulations. Our steps will proceed as follows:
\begin{enumerate}
\item We present in section \ref{sec:numerical simulation} numerical simulations to better describe the interaction we are to analyse. Simulations also motivate two important tools for asymptotic analysis -- multiple scale ansatz and fast-slow decomposition. 
\item Motivated by simulations,  we present in section \ref{sec:WKB equation results} our multiple scale model to predict $\epsilon V^{R}$, "precursor" radiation of $v$. Connections to its numerical motivations (section \ref{sec:numerical simulation}) are indispensably focused. 

\item Also motivated by simulations, we mathematically clarify in section \ref{sec:fast-slow decomposition} the fast-slow decomposition. Connections to their numerical motivations (section \ref{sec:numerical simulation}) are indispensably focused . This allows us to solve equations arisen in WKB theory by reducing PDEs to ODEs.

\item Equipped with the multiple-scale model and fast-slow decomposition, we are able, in section \ref{sec:solutions in moving frame}, to predict the simple analytic solution for the "precursor" $v$-radiation $\epsilon V^{R}$ and explain the mechanism of interaction. It is convenient to transform equations into moving frame in this section.

\end{enumerate}

\subsection{Numerical simulation with $\epsilon =10^{-2}$}
\label{sec:numerical simulation}

\begin{figure}[tb]
\centerline{
\includegraphics[width=\textwidth]{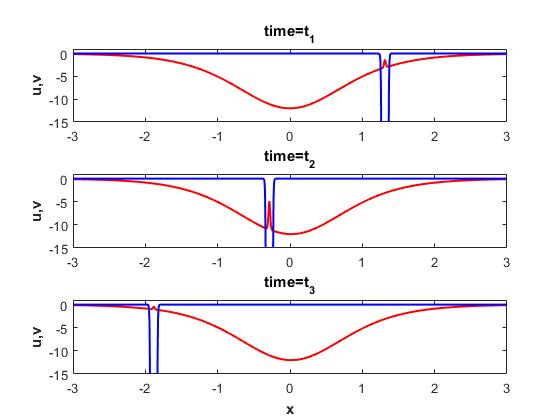}
}
\caption{Observation of $U^{F}(x,X,T)$, a fleeting solitary structure inside $u(x,X,T)$. Slow and red one is \textcolor{red}{$u$ wave} while the fast, sharp and blue one is \textcolor{blue}{$v$ wave}.}
\label{fig:observation of U1}	
\end{figure}

\begin{figure}[tb]
\centerline{
\includegraphics[width=\textwidth]{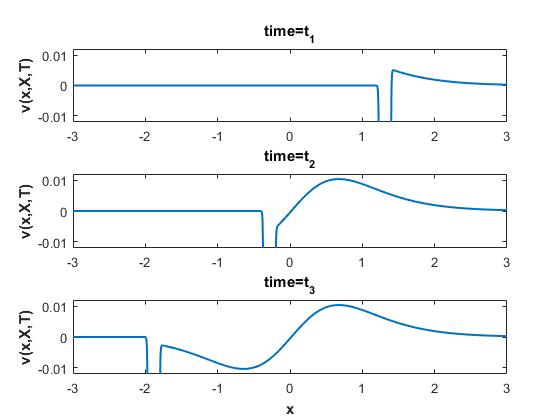}
}
\caption{Observation of $\epsilon V^{R}(x,X,T)$ as precursor $v$ radiation; observed from $v(x,X,T)$. Comparatively, figure \ref{fig:V2, order 1 of u,v}
 is a zoom-in in order to observe $V^{R}(x,X,T)$.}
\label{fig:observation of V2 descry}	
\end{figure}

As for ansatz (\ref{eq:ansatz}), $V^{0}(X,T)= K(X-cT)$ is approximately the original sharp soliton of wave $v$; correspondingly, $U^{0}$ is the base of wave $u$, approximately $U^{0} = K(x)$ since
$$K(x-c\epsilon^3 T)  = K(x)-c\epsilon^3 K_x(x)T = K(x)+O(\epsilon^3).$$

 At first glimpse, fleeting solitary structure 
\begin{equation}
U^{F}(x,X,T) = u(x,X,T)-U^{0}(x)+o(\epsilon^0)= u(x,X,T)-K(x)+o(\epsilon^0) 
\label{eq:U fleet from simulation}
\end{equation} is very eye-catching in figure \ref{fig:observation of U1}; after a zoom-in, slow--moving "precursor" radiation $\epsilon V^{R}$ (figure \ref{fig:observation of V2 descry},\ref{fig:V2, order 1 of u,v}) is a more intriguing feature. Although less eye-catching,  observations of $U^{F}$ (figure \ref{fig:snapshots of U1 descry}) is naturally presented later on. Afterwards, $U^{B}$ (figure \ref{fig:snapshots of U1 scrutine}) is more obscure (overshadowed by $U^{0}$) but validate our multiple scale model while establishing our model. Lastly, observations of $V^{F}$ is difficult and actually do not play a major role while establishing our model but help us validate our model. 

 Under this philosophy, we next orderly present observation of $U^{F},V^{R}, U^{B}$ in this section (section \ref{sec:numerical simulation}).

\subsubsection{Observation of $U^{F}$}
\label{subsubsection:observation of U1}
\begin{figure}[tb]
\centerline{
\includegraphics[width=\textwidth]{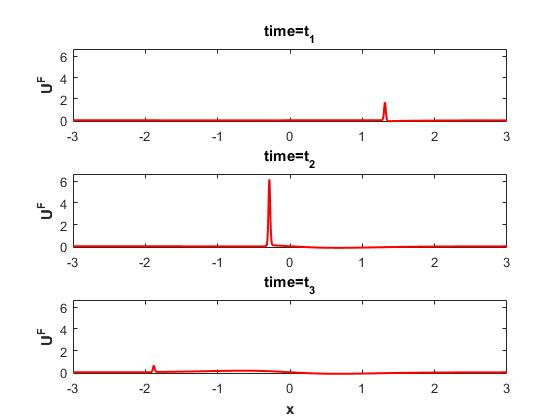}
}
\caption{Snapshots of $U^{F}(x,X,T)
$ in simulation with $\epsilon=10^{-2}$. See \ref{app:processing of simulation data to plot figures} for details of data processing.
}
\label{fig:snapshots of U1 descry}
\end{figure}

 Fleeting solitary structure $U^{F}$ in figure \ref{fig:observation of U1} is very eye-catching; it is numerically evaluated by (\ref{eq:U fleet from simulation}) and is plotted in figure \ref{fig:snapshots of U1 descry}.

\begin{figure}[tb]
\centerline{
\includegraphics[width=\textwidth]{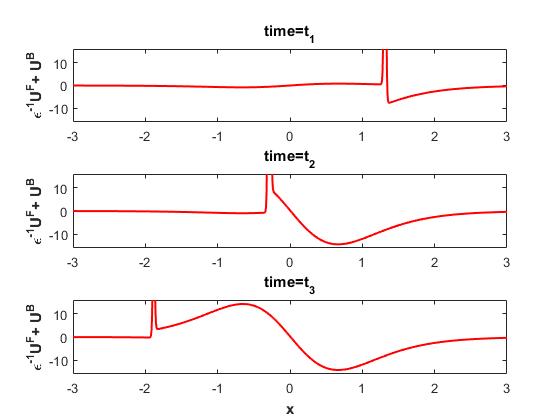}
}
\caption{Observation of $R(x)$ shape and Heaviside like structure $H^{U}(X-cT)$ from $\left( \epsilon^{-1}U^{F}+U^{B}\right)(x,X,T)
$ in simulation with $\epsilon=10^{-2}$. See \ref{app:processing of simulation data to plot figures} for details of data processing.
}
\label{fig:snapshots of U1 scrutine}
\end{figure}

From the simulation, 
\begin{enumerate}
\item $
\max_{x,X,T\in \BBR}U^{F}\simeq 6.57\sim 6.87,
$
for $\epsilon=10^{-2},1.4\times 10^{-2},2\times 10^{-2},2.5\times 10^{-2}$. In other words, $U^{F}$ appears to be a fleeting solitary structure in figure \ref{fig:snapshots of U1 descry} such that $\dps \max_{x,X,T\in \BBR}U^{F}(x,X,T)$ has nothing to do with $\epsilon$.
\item As time $T$ goes, height of $U^{F}$ seems proportional to $U^{0} = K(x) + O(\epsilon^3)$.

	We prefer saying $U^{F} \propto \sech^2(x)$ instead since $U^{0} < 0 $ and $\max | U^{0}| = 12$. Such preference (together with normalization of $R(x)$ later on) is very reader friendly since it , to some extent, ensures data consistency
\begin{itemize}
\item in figure \ref{fig:observation of V2 descry},\ref{fig:V2, order 1 of u,v},\ref{fig:heaviside structure H2}; table \ref{tab:theoretical maximum V1 V2}, representing size of $U^{B}$;
\item in figure \ref{fig:snapshots of U1 scrutine},\ref{fig:heaviside structure G2}; first part of figure \ref{fig:G2 H2}; table \ref{tab:theoretical maximum U1 U2}; derivation in (\ref{eq:right boundary value for G2 use integral of G}), representing size of $V^{R}$.
\end{itemize}
\end{enumerate}

\begin{figure}[tb]
\centerline{
\includegraphics[width=\textwidth]{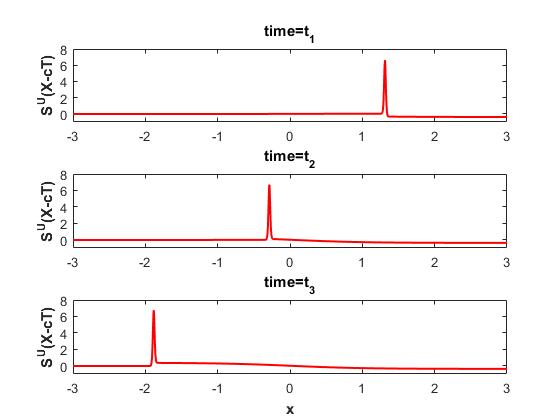}
}
\caption{Observation of solitary $S^{U}(x,X,T)$ from snapshots of $U^{F}(x,X,T)/\sech^2(x)$ in simulation with $\epsilon=10^{-2}$. See \ref{app:processing of simulation data to plot figures} for details of data processing.
}
\label{fig:snapshots of G}
\end{figure}

 This feature inspires us to plot $ U^{F}(x,X,T)/ \sech^2(x)$ in figure \ref{fig:snapshots of G}.
  Notice the dominant part of $U^{F}(x,X,T) / \sech^2(x)$ in figure \ref{fig:snapshots of G} seems to be a stable fast-moving solitary structure $S^{U}(X-cT)$, we introduce $S^{U}(X-cT)$. 

To conclude, we suppose that $U^{F}(x,X,T)$ has an exact solution in the form:
\begin{displaymath}
U^{F}(x,X,T) = \sech^2(x)S^{U}(X-cT).
\end{displaymath}
%

\subsubsection{Observation of $\epsilon V^{R}$}
\begin{figure}[tb]
\centerline{
\includegraphics[width=\textwidth]{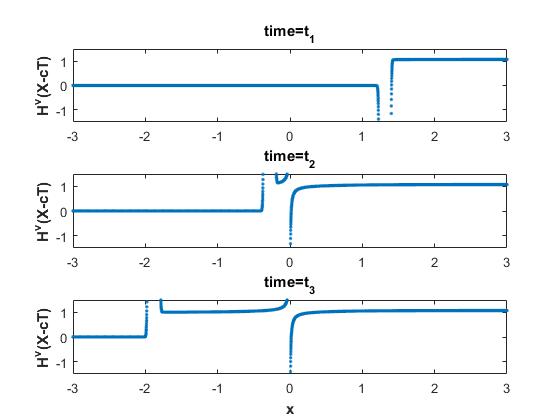}%
}
\caption{Heaviside-like structure $H^{V}(x-cT)$ from $ V^{R}(x,X,T)/R(x)$ in simulation with $\epsilon =10^{-2}$. See \ref{app:processing of simulation data to plot figures} for details of data processing.}
\label{fig:heaviside structure H2}
\end{figure}

As one of goals of the paper, "precursor" radiation \cite{biello2009nonlinearly} in $v$ is actually term $\epsilon V^{R}(x,X,T)$ at the bottom of $v(x,X,T)$.  Figure \ref{fig:V2, order 1 of u,v} is a close-up of $v$ near $x=0$ and implies that "precursor" radiation is of order $O(\epsilon^1)$ and $V^{R}$ looks like multiplication of 
\begin{enumerate}
\item  a fast-moving Heaviside--like structure $H^{V}(X-cT)$ with $\lim_{\xi\to-\infty}H^{\mbox{fast}}(\xi) =0$ and $\lim_{\xi\to \infty}H^{\mbox{fast}}(\xi) = \max_{x,X}V^{R}(x,X,\infty)$ where 
$V^{R}(x,X,\infty) = \lim_{T\to \infty}V^{R}(x,X,T)$. It provides an effect that precursor $v$ radiation seems scanned out by this fast Heaviside-like structure in figure \ref{fig:V2, order 1 of u,v}.
\item and a (normalized) centrosymmetric object $R(x) \propto K_x(x)$:
\begin{equation} 
R(x) \doteq \frac{\sinh(x)\cosh^3(x)}{\max_x \sinh(x)\cosh^3(x) } = \frac{3\sqrt{3} }{2}\sinh(x)\cosh^3(x). \label{eq:radiation centrosymmetric object}
\end{equation}

Notice we let $\max_x R(x) = -\min_x R(x) =1 $ so that $C^{R}_V$ in (\ref{eq:conclusion of precursor v radiation in the introduction}) can represent height of $V^{R}$. Mentioned again, such normalization of $R(x)$ (together with preference of $\sech^2(x)$ over $K(x)$) is reader friendly since it, to some extent, ensures data consistency
\begin{itemize}
\item in figure \ref{fig:observation of V2 descry},\ref{fig:V2, order 1 of u,v},\ref{fig:heaviside structure H2}; table \ref{tab:theoretical maximum V1 V2}, representing size of $U^{B}$;
\item in figure \ref{fig:snapshots of U1 scrutine},\ref{fig:heaviside structure G2}; first part of figure \ref{fig:G2 H2}; table \ref{tab:theoretical maximum U1 U2}; derivation in (\ref{eq:right boundary value for G2 use integral of G}), representing size of $V^{R}$.
\end{itemize}
\end{enumerate}

\begin{figure}[tb]
\centerline{
\includegraphics[width=\textwidth]{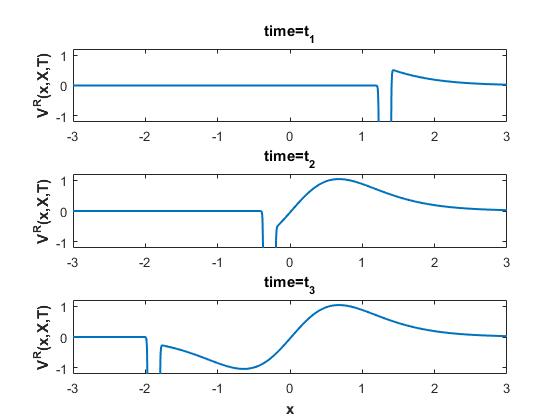}
}
\caption{Observation of $R(x)$ shape and Heaviside like structure $H^{U}(X-cT)$ from $V^{R}(x,X,T)$ and observed directly at the bottom of $\epsilon^{-1}v(x,X,T)$. See \ref{app:processing of simulation data to plot figures} for details of data processing. A zoom-in of figure \ref{fig:observation of V2 descry}.}
\label{fig:V2, order 1 of u,v}
\end{figure}

This feature inspires us to plot $V^{R}(x,X,T)/R(x)$ in figure \ref{fig:heaviside structure H2} and we basically only expect it to be "Heaviside-like". Indeed, a Heaviside-like structure  appears with $\lim_{\xi\to -\infty}H^{V}(\xi) =0$ and $\lim_{\xi\to \infty}H^{V}(\xi) = \max_{x,X} V^{R}(x,X,\infty)$ (despite the (numerical) singularity of $\csch x$, near the origin $x=0$).

To conclude, we suppose that $V^{R}(x,X,T)$ has an exact solution in the form:
\begin{displaymath} 
V^{R}(x,X,T)=R(x) H^{V}(X-cT).
\end{displaymath}

\subsubsection{Observation of $\epsilon U^{B}$}
After noticing that small bruise appear at the bottom of $u$ in figure \ref{fig:snapshots of U1 descry}, a scrutiny reveals structure $\epsilon U^{B}$ of order $O(\epsilon^1)$ and we zoom in figure \ref{fig:snapshots of U1 scrutine}:
$U^{B}(x,X,T)$ looks like $R(x)$ in (\ref{eq:radiation centrosymmetric object}) multiplied by a Heaviside--like structure $H^{U}(X-cT)$. 

\begin{figure}[tb]
\centerline{
\includegraphics[width=\textwidth]{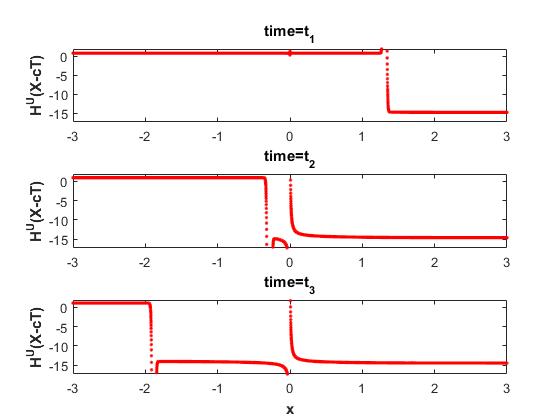}
}
\caption{Heaviside-like structure $H^{U}(X-cT)$ from $U^{B}(x,X,T)/R(x)$ in the simulation with $\epsilon =10^{-2}$. See \ref{app:processing of simulation data to plot figures} for details of data processing.}
\label{fig:heaviside structure G2}
\end{figure}

This feature inspires us to plot $U^{B}(x,X,T)/R(x)$ figure \ref{fig:heaviside structure G2} and we basically only expect $H^{U}(X-cT)$ to be "Heaviside--like".  Indeed, despite (numerical) singularity of $\csch x$, near the origin $ x=0$， a Heaviside-like structure appears with $\lim_{\xi\to -\infty}H^{U}(\xi) =0$ and $\lim_{\xi\to \infty}H^{U}(\xi) = \max_{x,X} U^{B}(x,X,\infty)$.

To conclude, we have shown that $U^{B}(x,X,T)$ has an exact solution in the form:
\begin{displaymath}
U^{B}(x,X,T)= R(x) H^{U}(X-cT).
\end{displaymath}

\section{Asymptotic approximation of the coupled KdV}

We will first perform all of numerical simulations in the case with $\epsilon =10^{-2}$. The essence of simulation is captured by this example, which can easily be generalized to the case with $\epsilon \to 0$.  

These simulations also validate multiple-scale ansatz (\ref{eq:ansatz}) and the fast-slow decomposition (\ref{eq:structure of U1}, \ref{eq:structure of V1}, \ref{eq:structure of U2}, \ref{eq:structure of V2}) in case of $\epsilon\to 0$.

\subsection{Multiple scale model: ansatz and equations}
\label{sec:WKB equation results}
After concluding ansatz (\ref{eq:ansatz}) from numerical simulations of (\ref{eq:nonlinearly coupled KdV equation of Biello}), we substitute the ansatz (\ref{eq:ansatz}) as well as $\frac{\partial }{\partial x}\to \epsilon^{-1}\frac{\partial }{\partial X}+\frac{\partial }{\partial x}$,
$\frac{\partial }{\partial t} \to \epsilon^{-3}\frac{\partial }{\partial T}$ into Biello's equations (\ref{eq:nonlinearly coupled KdV equation of Biello}) and get (\ref{eq:nonlinearly coupled KdV equation of Biello}) with respect to different order of $\epsilon$ as:

\numparts
\begin{eqnarray}
&& \epsilon^{-3}\cL\left(U^{F}\right)+\epsilon^{-2}[\cL\left(U^{B}\right)-3U^{F}_{xXX}] \nonumber\\
=&& \frac12 \left\{\epsilon^{-3}\left[(V^{0}U^{F})_X +U_{0}V_{0,X}\right]
+\epsilon^{-2}\left[(V^{0}U^{B})_X+(U_{0,x}+U^{F}_x)V^{0}\right]\right\},
\label{eq:WKB U}
\end{eqnarray}
\begin{eqnarray}
&& \epsilon^{-5}\cK(V^{0})+\epsilon^{-3}\left[\cL\left(V^{F}\right)+(V^{0}V^{F})_X\right]\nonumber \\
&& +\epsilon^{-2}[\cL(V^{R}) +V^{0}V^{F}_x+(V^{R}V^{0})_X-3V^{F}_{xXX}] \nonumber \\
=&&  \frac12 \left\{\epsilon^{-3}\left[(V^{0}U^{F})_X+U_{0}V_{0,X}\right] 
+\epsilon^{-2}\left[(V^{0}U^{B})_X+(U_{0,x}+U^{F}_x)V^{0}\right]\right\}.\label{eq:WKB V}
\end{eqnarray}
\endnumparts
where 
\begin{displaymath}
\cL(f)=f_{T}-f_{XXX},\qquad \cK(f)=\cL(f)+ff_X.
\end{displaymath}

 From (\ref{eq:WKB V}), we can notice that  $U^{F}, V^{F}, U^{B}$ are indispensable for solving the equation for $V^{R}$ although "precursor" radiation $V^{R}$ is our focus of attention in the simulation.

\subsection{Fast-slow decomposition}
\label{sec:fast-slow decomposition}
Fast-slow decomposition basically means we decompose functions with respect to two pairs of time-length scales -- a small length scale $X$ with fast time $T$, and a large length scale $x$. At first glimpse, fast-slow decomposition just seems to be an outcome of multiple scale analysis. However, the decomposition is particularly listed since
\begin{enumerate}
\item together with "Heaviside-like" structures $H^{U},H^{V}$, it plays a crucial role to validate the ansatz;
\item it plays an important role in reducing PDEs for $U^{F},V^{F},U^{B},V^{R}$ into ODEs (\ref{eq:ode for G}, \ref{eq:ode for H}, \ref{eq:ode for G2}, \ref{eq:ode for H2}) in section \ref{sec:solutions in moving frame}.
\end{enumerate}

Notice that in the stationary frame of reference, the $V^{0}$ soliton does not depend on $\tau$. The equation (\ref{eq:U1 moving frame}) for $U^{F}$ comes endowed with boundary values $U^{F}(x,\xi,\tau)\to 0$ as $\xi \to \pm\infty$ and initial condition $U^{F}(x,\xi,T) \to 0$ as $\tau\rightarrow -\infty$. 

Eventually, the decomposition turn to be:
\numparts
\begin{equation}U^{F}(x,X,T)=\sech^2(x)S^{U}(X-cT),\label{eq:structure of U1}\end{equation}
\begin{equation}V^{F}(x,X,T)=\sech^2(x)S^{V}(X-cT), \label{eq:structure of V1}\end{equation}
\begin{equation} U^{B}(x,X,T)=R(x)H^{U}(X-cT), \label{eq:structure of U2}\end{equation}
\begin{equation}
V^{R}(x,X,T)=R(x) H^{V}(X-cT).  \label{eq:structure of V2}
\end{equation}
\endnumparts
or for short:
\begin{eqnarray*}
U^{F}(x,\xi,\tau)=\sech^2(x) S^{U}(\xi), && \qquad V^{F}(x,\xi,\tau)=\sech^2 (x) S^{V}(\xi),\\
 U^{B}(x,\xi,\tau)=R (x)H^{U}(\xi), && \qquad V^{R}(x,\xi,\tau)=R(x) H^{V}(\xi),
\end{eqnarray*}
since $U_{0}$ component of wave $u$ almost unmoved with respect to wave $v$. 
Here moving frame $\xi=X-cT$ is presented in advance and will be utilized as a customized frame while reducing PDEs into ODEs in section \ref{sec:solutions in moving frame}.

\subsection{Solving for Biello's equation in moving frame}
\label{sec:solutions in moving frame}
Substituting ansatz (\ref{eq:ansatz}) into (\ref{eq:nonlinearly coupled KdV equation of Biello}), Biello's equations become (\ref{eq:WKB U}, \ref{eq:WKB V}), which can be written out versus different order of $\epsilon$ as:
\numparts
\begin{equation}
\cL(U^{F})- \frac{\left[V^{0}U^{F}\right]_X}2  \equiv \frac{U_{0}V_{0,X}}2, \label{eq:U1 inhomogeneous term at right hand side}
\end{equation}
\begin{equation}
\cL(V^{F})+\left[V^{0}V^{F}\right]_X   \equiv \frac{\left[U^{F}V^{0}\right]_X+U_{0}V_{0,X} }2, \textbf{\label{eq:V1 inhomogeneous term at right hand side}}
\end{equation}
\begin{equation}
\cL(U^{B})- \left[V^{0}U^{B}\right]_X \equiv \left[U_{0,x}+U^{F}_x\right]V^{0} + 3U^{F}_{xXX}, \label{eq:U2 inhomogeneous term at right hand side}
\end{equation}
\begin{equation}
\cL(V^{R}) +\left[V^{0} V^{R}\right]_X  \equiv \left[V^{0}U^{B}\right]_X +\left[U_{0,x}+U^{F}_x\right]V^{0}- V^{0}V^{F}_x + 3V^{F}_{xXX}. \label{eq:V2 inhomogeneous term at right hand side}
\end{equation}
\endnumparts
where 
\begin{equation}
\cL(f)=f_{T}-f_{XXX}, \qquad U_{0}=K(x),\qquad V^{0}=K(X-cT).
\label{eq:operators just for convenience}
\end{equation}

These equations have simple structures, three aspects of which are very illuminating:
\begin{enumerate}
\item They each have an in-homogeneity listed, separately, on right hand sides of each equation for $U^{F},V^{F},U^{B},V^{R}$.
\item The first order $T$-derivatives and the third order $X$-derivatives are endowed by linear KdV operator $L$.
\item The in-homogeneous terms for $V^{F},U^{B}$ only involves $U^{F}$; the one for $V^{R}$ involves $U^{F},V^{F}$. Therefore, we can solve equations $V^{F},U^{B}$ after solving $U^{F}$ and eventually solve $V^{R}$.
\end{enumerate}

\subsubsection{Perspectives in moving frame}
In the limit that $\epsilon\to 0$, it is tantamount to viewing equations in moving frame $\xi(X,T)\equiv X-cT$, $\tau(X,T) \equiv T $ which is naturally inspired by numerical simulations by referring to figure \ref{fig:snapshots of U1 descry}, \ref{fig:snapshots of U1 scrutine}, \ref{fig:V2, order 1 of u,v} for $U^{F},U^{B},V^{R}$ and correspondingly, figure \ref{fig:snapshots of G}, \ref{fig:heaviside structure G2}, \ref{fig:heaviside structure H2} for $S^{U},H^{U},H^{V}$. Chain rule implies
\begin{eqnarray*}
\left[
\frac{\partial}{\partial X}, \frac{\partial}{\partial T}\right] 
&&  \equiv  \left[
\frac{\partial}{\partial \xi}, \frac{\partial}{\partial \tau}\right] 
\left[\begin{array}{cc}
\frac{\partial \xi}{\partial X}(X,T) &　\frac{\partial \xi}{\partial T}(X,T)  \\
\frac{\partial \tau}{\partial X}(X,T) &　\frac{\partial \tau}{\partial T}(X,T) 
\end{array}
\right] \equiv   \left[
\frac{\partial}{\partial \xi}, \frac{\partial}{\partial \tau}\right] 
 \left[\begin{array}{cc}
1 &　4 \\
0 & 1
\end{array}\right] \\
\Leftrightarrow   && 
 \frac{\partial}{\partial X}\equiv \frac{\partial}{\partial \tau}-c\frac{\partial }{\partial \xi},  \qquad \frac{\partial }{\partial T}\equiv \frac{\partial}{\partial \tau},
\end{eqnarray*}
or for short, 
\begin{eqnarray}
&& \xi(X,T)\equiv X-cT,  \qquad \tau(X,T) \equiv T \nonumber \\
 \Rightarrow  && \frac{\partial}{\partial X}\equiv \frac{\partial}{\partial \tau}-c\frac{\partial }{\partial \xi},  \qquad \frac{\partial }{\partial T}\equiv \frac{\partial}{\partial \tau}.
\label{eq:chain rule derivative}
\end{eqnarray}

Performing the changing of variables on equations (\ref{eq:U1 inhomogeneous term at right hand side},\ref{eq:V1 inhomogeneous term at right hand side},\ref{eq:U2 inhomogeneous term at right hand side},\ref{eq:V2 inhomogeneous term at right hand side})
and substituting $U_{0}\equiv K(x),V^{0} \equiv K(\xi)$ by (\ref{eq:operators just for convenience}) incidentally. We find (\ref{eq:U1 inhomogeneous term at right hand side}) for $U^{F}$ is a linear, in-homogeneous PDE independent of $V^{F},U^{B},V^{R}$,
\numparts
\begin{equation}
\cL(U^{F}) -\frac{\left[ K(\xi)U^{F}\right]_{\xi}}{2} \equiv \frac{ K(x)K(\xi)_{\xi}}{2},
\label{eq:U1 moving frame}
\end{equation}
where linear KdV operator turns into
$
\cL(f)\equiv f_{\tau}-cf_{\xi}-f_{\xi\xi\xi}.
$ by chain rule (\ref{eq:chain rule derivative}).

(\ref{eq:V1 inhomogeneous term at right hand side},\ref{eq:U2 inhomogeneous term at right hand side}) for $V^{F},U^{B}$ are also linear, each with an in-homogeneity which depends on $U^{F}$,

\begin{equation}
\cL(V^{F})+\left[K(\xi)V^{F}\right]_{\xi}\equiv \frac{K(x)U^{F}_{\xi}+K(x)K(\xi)_{\xi} }2,
\label{eq:V1 moving frame}
\end{equation}
\begin{equation}
\cL(U^{B})- \left[K(\xi)U^{B}\right]_{\xi} \equiv \left[F_x(x)+U^{F}_x\right]K(\xi) + 3U^{F}_{x\xi\xi}.
\label{eq:U2 moving frame}
\end{equation}

Finally, (\ref{eq:V2 inhomogeneous term at right hand side}) for $V^{R}$ is also linear, with an in-homogeneity which depends on $U^{F},V^{F},U^{B}$, 
\begin{eqnarray}
&& \cL(V^{R}) +\left[K(\xi)V^{R}\right]_{\xi} \nonumber\\
 \equiv && \left[K(\xi)U^{B}\right]_{\xi}+\left[F_x(x)+U^{F}_x\right]K(\xi)- K(\xi)V^{F}_x + 3V^{F}_{x\xi\xi}. 
\label{eq:V2 moving frame}
\end{eqnarray}
\endnumparts

Under this philosophy, in order to predict $V^{R}$, we need to first orderly predict $U^{F}$, $V^{F}$, $U^{B}$.  Later in this section (section \ref{sec:solutions in moving frame}), we will predict them in order of $U^{F}$, $V^{F}$, $U^{B},V^{R}$.

\subsubsection{Solution of $U^{F}$}
 From fast-slow decomposition, we know that  (\ref{eq:U1 moving frame}) has an exact solution in the form:
\begin{equation}
U^{F}(x,\xi,\tau) = \sech^2 (x) S^{U}(\xi),
\label{eq:fast-slow decom U1}
\end{equation}
 where we actually seek the solution for $U^{F}$ independent of $\tau$.

 Substituting (\ref{eq:fast-slow decom U1}) into (\ref{eq:U1 moving frame}) we find the ODE for $G$:
\begin{equation}
S^{U}_{\xi\xi}-\left[4+6\sech^2(\xi)\right]S^{U}=-72\sech^2(\xi).
\label{eq:ode for G}
\end{equation}

\subsubsection{Solution of $V^{F}$}
Again according to fast-slow decomposition, we seek a solution of the form:
\begin{equation}
V^{F}(x,\xi,\tau) = \sech^2 (x) S^{V}(\xi),
\label{eq:fast-slow decom V1}
\end{equation}

After substituting (\ref{eq:fast-slow decom U1} ,\ref{eq:fast-slow decom V1}) into equation (\ref{eq:V1 moving frame}), we find the ODE for $S^{V} (\xi)$

\begin{equation}
S^{V}_{\xi\xi}-[4-12\sech^2(\xi)]S^{V}=6[S^{U}(\xi)-12]\sech^2(\xi).
\label{eq:ode for H}
\end{equation}
with boundary values $\dps \lim_{\xi\to-\infty}S^{V}(\xi)=\lim_{\xi\to\infty} S^{V}(\xi)=0$ according to figure \ref{fig:snapshots of G} of numerical simulation. Figure \ref{fig:H at xi=0.0 and the one from WKB} (second) plots $S^{V}$ with $S^{V}(-10)=S^{V}(10)=0$.
\subsubsection{Solution of $U^{B}$}

Again according to fast-slow decomposition, we seek a solution of the form:
\begin{equation}
U^{B}(x,\xi,\tau) = R(x) H^{U}(\xi).
\label{eq:fast-slow decom U2}
\end{equation}

After substituting (\ref{eq:fast-slow decom U1} ,\ref{eq:fast-slow decom V1}, \ref{eq:fast-slow decom U2}) into equation (\ref{eq:U2 moving frame}), we find that the ODE for $H^{U}(\xi)$

\begin{equation}
H^{U}_{\xi\xi}-[4+6\sech^2(\xi)]H^{U}=\frac{4\sqrt{3}}3\left[2\int_{-\infty}^{\xi}[12-S^{U}(\eta)]\sech^2(\eta)d\eta+S^{U}_{\xi}\right].
\label{eq:ode for G2}
\end{equation}
with left boundary value $\dps \lim_{\xi\to-\infty}H^{U}(\xi) = 0$ according to figure \ref{fig:heaviside structure G2} from numerical simulation; by taking $\dps \lim_{\xi\rightarrow\infty}$ on both sides of (\ref{eq:ode for G2}), right boundary value of appears as $H^{U}(10)=-\frac{4\sqrt{3}}{9} \int_{\BBR}S^{U}(\eta)d\eta = -15.37$ (see \ref{app:Derivation of right boundary values for HU HV} for detail). To conclude, figure \ref{fig:G2 H2} (first) shows Heaviside-like $H^{U}$.
\subsubsection{Solution of $V^{R}$}
Finally,  we seek a solution of the form for "precursor" radiation $V^{R}$:
\begin{equation}
V^{R}(x,\xi,\tau) = R (x)H^{V}(\xi).
\label{eq:fast-slow decom V2}
\end{equation}

After substituting (\ref{eq:fast-slow decom U1}, \ref{eq:fast-slow decom V1}, \ref{eq:fast-slow decom U2}, \ref{eq:fast-slow decom V2}) into equation (\ref{eq:V2 moving frame}), we find that the terms give an equation for $H^{V}(\xi)$

\begin{eqnarray}
&& H^{V}_{\xi\xi}-[4-12\sech^2(\xi)]H^{V} = \frac{4\sqrt{3}}3 \cdot \nonumber \\
 && \left[H^{U}(\xi)\sech^2(\xi)+S^{V}_{\xi} 
+ 2\int_{-\infty}^{\xi}\left[12-S^{U}(\eta)+2S^{V}(\eta)\right]\sech^2(\eta)d\eta\right]. \label{eq:ode for H2}
\end{eqnarray} 
with left boundary value $\dps \lim_{\xi\to-\infty}H^{V}(\xi)= 0$ according to figure \ref{fig:heaviside structure H2} from numerical simulation; by taking $\dps \lim_{\xi\rightarrow\infty}$ on both sides of (\ref{eq:ode for H2}), right boundary value appears as $H^{V}(10)=-\frac{4\sqrt{3}}9\int_{\BBR}S^{V}(\eta)d\eta\simeq 4.11$ (see \ref{app:Derivation of right boundary values for HU HV} for detail). To conclude, figure \ref{fig:G2 H2} (second) shows Heaviside-like $H^{V}$.

\subsection{Conclusion of solutions}

With our multiple-scale model, we can predict $U^{F},V^{F},U^{B},V^{R}$ in ansatz (\ref{eq:ansatz}). Details of analytic solution utilizes fast-slow decomposition (\ref{eq:structure of U1}, \ref{eq:structure of V1}, \ref{eq:structure of U2}, \ref{eq:structure of V2}) and solve fast-moving part of them via ODEs for $S^{U},S^{V},H^{U},H^{V}$.

As a special case, $\epsilon V^{R}$, the \textbf{energy exchange} and this subtle \textbf{"precursor" radiation} can be analytically predicted. Our multiple scale model predicts (\ref{eq:conclusion of precursor v radiation in the introduction}):
$$\epsilon V^{R}= \epsilon C^{R}_{V} R(x),$$
as $T\to \infty$ where $C^{R}_{V} = \lim_{\xi\to\infty}H^{V}(\xi) = 4.11$. 

Comparisons between model predictions and numerical simulation (with various $\epsilon$) are summarized in section \ref{app:comparison of sizes between WKB theory}.

\section{Comparison to Numerical Simulations}
\label{app:comparison of sizes between WKB theory}
We compare our model predictions to numerical simulations of the Biello's equation (\ref{eq:nonlinearly coupled KdV equation of Biello}). The system is solved in a doubly periodic domain using a pseudo--spectral method. These simulations were performed with $\triangle x = \frac{25}{N}$ for $N=2^{15}$, with $\epsilon= 0.1, 0.025, 0.02, 0.014, 0.01$, where $\frac{\triangle t}{(\triangle x)^3}\simeq 1 $
. All solutions were monitored for conservation of energy, Hamiltonian and (two) mean fields to a relative accuracy of at least $10^{−10}$. The last nonlinear energy-conserved term is also an important feature in wave turbulence \cite{krechetnikov2009origin}. Interestingly \cite{vodova2015symmetries} points out our Biello-Majda system has only these four conserved quantities.

\begin{table}
\caption{Maximum of $U^{F}$ and second local extrema of $U^{B}$: between model predictions and numerical simulations ($^{BI}$: data before interaction; $^{DI}$: data during interaction).

\begin{itemize}
\item numerical simulations: when it comes to $\max U^{F}$, we are evaluating maximum of solitary structure shown in figure \ref{fig:snapshots of U1 descry}; 

speaking of second local extrema of $U^{B}$, we refer to the local minima on the right of figure \ref{fig:snapshots of U1 scrutine};
\item Predicted values: when it comes to predicted $\max U^{F}$, predicted $U^{F}$ is from fast-slow decomposition (\ref{eq:fast-slow decom U1}) where  $S^{U}$ from (\ref{eq:ode for G}) is numerically presented via finite difference method with boundary values $\dps \lim_{\xi\to \pm\infty}S^{U}(\xi) = 0$  (see figure \ref{fig:Schwartz solution G1} for consistency of $S^{U}$);

speaking of second local extrema of $U^{B}$, we refer to the local minima of $U^{B}$ from fast slow decomposition (\ref{eq:fast-slow decom U2}). $H^{U}$ from (\ref{eq:ode for G2}) is numerically presented via finite difference method with boundary values $H^{U}(-10)=0, H^{U}(10)=-\frac{4\sqrt{3}}9\int_{\BBR}S^{U}(\eta)d\eta\simeq -15.37$ (see figure \ref{fig:G2 H2}).
\end{itemize} 
}
\begin{indented}
\item[]
\begin{tabular}{@{}lll}
\br
$\epsilon$ & $\max U^{F}$  & second local extrema of $U^{B}$ \\ 
\mr prediction & $6.68$ & $-15.37$ \\
\mr
$0.1$ 
 &$6.66 \sim 6.87$   &  \\
\mr
$0.0025$
 &  $6.57  \sim 6.67^{BI}$  & $ -16.02 \sim -16.06 $ \\ 
\mr
$0.002$ 
 &  $6.69  \sim 6.71$ & $ -15.90 \sim -15.91$ \\ 
 \mr
$0.0014$
&  $6.69 \sim 6.70$  &$ -15.23 \sim -15.80$   \\ 
\mr
$0.001$
&  $6.62 \sim 6.70$  & $ -15.31 \sim -15.66 $ \\
\br
\end{tabular}
\end{indented}
\label{tab:theoretical maximum U1 U2}
\end{table}

In table \ref{tab:theoretical maximum U1 U2}, we show comparison of $\max U^{F}$ and second local extrema of $U^{B}$ between  predicted values numerical simulations. We see excellent agreement for $U^{F}$, $U^{B}$ between the model predictions and the numerical solutions.

\begin{table}
\caption{Maximum of $V^{F}$ and second local extrema of $V^{R}$: model prediction and numerical simulations ($^{BI}$: data before interaction; $^{DI}$: data during interaction).

\begin{itemize}
\item numerical simulations: when it comes to $\max V^{F}$, we are evaluating minimum of solitary structure shown in section \ref{app:observation of H} where $\min V^{F} = \min S^{V}$ due to fast-slow decomposition (\ref{eq:fast-slow decom V1});  

speaking of second local extrema of $V^{R}$, we refer to the local maxima on the right of figure \ref{fig:V2, order 1 of u,v};
\item Predicted values: when it comes to predicted $\max V^{F}$, predicted $V^{F}$ is from fast-slow decomposition (\ref{eq:fast-slow decom V1}) where  $S^{V}$ from (\ref{eq:ode for H}) is numerically presented via finite difference method with boundary values $\lim_{\xi\to \pm\infty}S^{V}(\xi) = 0$  (see figure \ref{fig:H at xi=0.0 and the one from WKB});

speaking of second local extrema of $U^{B}$, we refer to the local maxima of $V^{R}$ from fast slow decomposition (\ref{eq:fast-slow decom V2}). $H^{V}$ from (\ref{eq:ode for H2}) is numerically presented via finite difference method with boundary values $H^{V}(-10)=0, H^{V}(10)=-\frac{4\sqrt{3}}9\int_{\BBR}S^{V}(\eta)d\eta= -4.11$ (see figure \ref{fig:G2 H2}).
\end{itemize} }
\begin{indented}
\item[]
\begin{tabular}{@{}lll} 
\br
$\epsilon$  & $\min V^{F}$  & second local extrema of $V^{R}$ \\ 
\mr prediction & $\textcolor{red}{-9.77}$ & $\textcolor{red}{4.11}$ \\
\mr
$0.75$  & $-6.57^{DI}$   &  hard to detect\\  
\mr
$0.1$ 
& $ -7.20  \sim -7.09^{BI}$   & $1.11  \sim  1.13$\\
 &$-6.83 \sim  -6.28^{DI}$  & \\
\mr
$0.0025$ 
 & $-6.06\sim -6.00^{BI}$& $ 1.12  \sim  1.14$\\ 
 &$-5.90 \sim  -5.86 ^{DI}$  & \\ 
 \mr
$0.002$ 
  & $-5.90  \sim -6.02$  &$1.11 \sim 1.13$\\ 
  \mr
$0.0014$ 
   & $-5.91  \sim -5.90^{BI}$  &$ 1.01  \sim  1.10$\\ 
 &$-5.77 \sim  -5.75^{DI}$ & \\ 
 \mr
$0.001$ 
& $-5.89 \sim  -5.86^{BI}$  & $1.04 \sim 1.10$\\ 
 &$-5.88 \sim  -5.86^{BI2}$ &
\\ 
 &$ -5.76 \sim  -5.74^{DI}$ & \\ 
\br
\end{tabular}
\end{indented}
\label{tab:theoretical maximum V1 V2}
\end{table}

\subsection{Comparison of $S^{U},H^{U},H^{V}$}

It is not just maximum of $U^{F}$, minimum of $V^{F}$, second local extrema of $U^{B}$, $V^{R}$ where we see good agreement between theory and simulation.

Due to fast-slow decomposition, instead of comparing $U^{F}, V^{F},U^{B},V^{R}$, we compare solitary $S^{U},S^{V}$, Heaviside-like $H^{U},H^{V}$ for the sake of convenience. This is not only because $S^{U},S^{V},H^{U},H^{V}$ can be viewed in moving frame $\xi=X-cT$, but $S^{U},S^{V},H^{U},H^{V}$ have much simpler structures as well - solitary $S^{U},S^{V}$ and Heaviside-like $H^{U},H^{V}$.

\begin{figure}[tb]
\centerline{
\includegraphics[width=\textwidth]{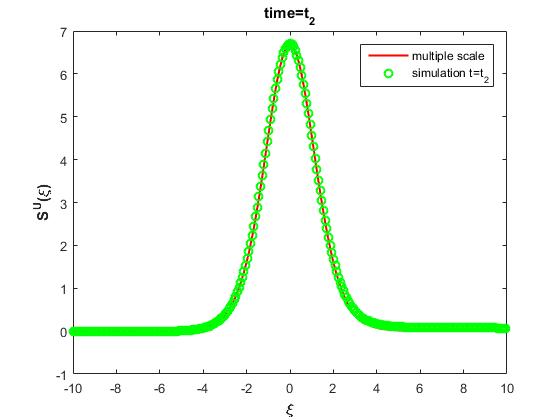}%
}
\caption{\textcolor{red}{$S^{U}(\xi)$} predicted by asymptotic model from ODE (\ref{eq:ode for G}) coincides with ones from \textcolor{green}{numerical simulation}: 

the predicted \textcolor{red}{$S^{U}(\xi)$} is plotted via finite difference method with Cauchy boundary condition $S^{U}(-10)=S^{U}(10)=0$.

\textcolor{green}{$S^{U}(\xi)$} from numerical simulation is plotted via $U^{F}(x,X,T)/\sech^2(x)$. See \ref{app:processing of simulation data to plot figures} for details of data processing.}
\label{fig:Schwartz solution G1}
\end{figure}

Figure \ref{fig:Schwartz solution G1} shows $S^{U}$
\begin{itemize}
\item predicted by asymptotic model from (\ref{eq:ode for G}) numerically via finite difference method with boundary condition $ \lim_{\xi\to \pm \infty}S^{U}= 0$ and 
\item the ones from simulations plotted via $U^{F}(x,X,T)/\sech^2(x)$ in section \ref{subsubsection:observation of U1} (same philosophy as figure \ref{fig:snapshots of G}).
\end{itemize}
The solitary structure is evident and our asymptotic model precisely predict solitary $U^{F}$ since two plots overlap each other.

\begin{figure}[tb]
\centerline{
\includegraphics[width=0.5\textwidth]{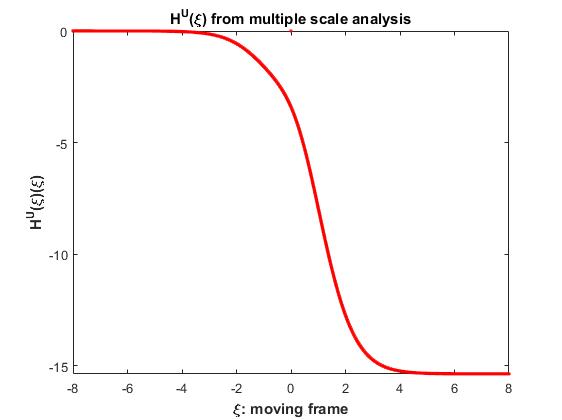}%
\includegraphics[width=0.5\textwidth]{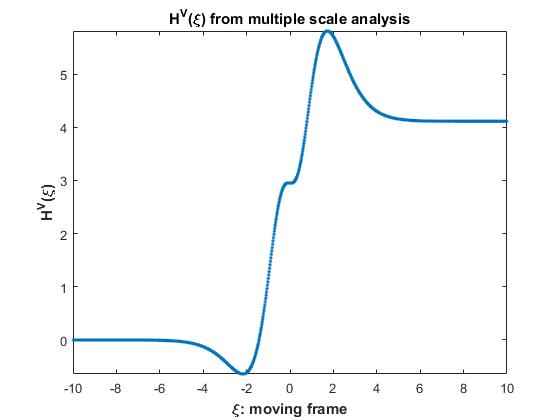}
}
\caption{First: $H^{U}(\xi)$ predicted by asymptotic model from (\ref{eq:ode for G2}) and numerically plotted via finite difference method with Cauchy boundary condition $\dps H^{U}(-10)=0, H^{U}(10)=-\frac{4\sqrt{3}}9\int_{\BBR}S^{U}(\eta)d\eta\simeq -15.37$; second: $H^{V}(\xi)$, predicted by asymptotic model from (\ref{eq:ode for H2}) and numerically plotted via finite difference method with Cauchy boundary condition $\dps H^{V}(-10)=0, H^{V}(10)=-\frac{4\sqrt{3}}9\int_{\BBR}S^{V}(\eta)d\eta\simeq 4.11$.}
\label{fig:G2 H2}
\end{figure}

Figure \ref{fig:G2 H2} (first) shows Heaviside-like $H^{U}$ predicted by asymptotic model from (\ref{eq:ode for G2}) numerically via finite difference method with boundary condition $\lim_{\xi\to -\infty}H^{U} = 0$ and $\lim_{\xi\to -\infty}H^{U} = -15.37$. This is in excellent agreement with simulation result in figure \ref{fig:snapshots of U1 scrutine},\ref{fig:heaviside structure G2} and therefore predict the exact size of $U^{B}$.

Finally, figure \ref{fig:G2 H2} (right) shows  Heaviside-like $H^{V}$ predicted by asymptotic model from (\ref{eq:ode for H2}) numerically via finite difference method with boundary condition $\lim_{\xi\to -\infty}H^{V} =0$ and $\lim_{\xi\to \infty}H^{V} \simeq 4.11$. Although this differs from asymptotic simulation shown in figure \ref{fig:observation of V2 descry},\ref{fig:V2, order 1 of u,v},\ref{fig:heaviside structure H2}, it is in agreement with the sign and relative size of $V^{R}, H^V$. Table \ref{tab:theoretical maximum V1 V2} concludes the difference.

\subsection{Observation and verification of $V^{F}, S^{V}$}

Numerical observation of $V^{F}$ is quite difficult once realizing $\epsilon^{-2}V^{0}$ part of wave $v$ is very sharp. On the other hand,we can validate our model by validating corresponding fast-slow decomposition (\ref{eq:structure of V1}).

 However, exact sizes of $V^{F},S^{V}$ are inconsistent between the model prediction and simulations. According to (\ref{eq:right boundary value for H2 use integral of H}), it is precisely integral of $S^{V}$ that determines right boundary value for $H^{V}$, which is the important feature for $H^{V}$. 
 
Details of this solitary structure $S^{V}$ (figure \ref{fig:H at xi=0.0 and the one from WKB}) with two small bumps at two sides is shown in {app:observation of H}.

\section{Conclusion}
\subsection{Theoretical contribution}
Our multiple scale model 
\begin{enumerate}
\item predict mechanism of interaction within this nonlinearly coupled KdV equations:
\begin{quote}
precursor $v$-radiation is formed firstly; then it interacts with $u$, forming $u$-radiation and $v$-radiation in-situ.
\end{quote}
\item predict with an analytic, asymptotic approximation, the "precursor" radiation generated during interaction of solitary waves in Biello's system.
\end{enumerate}

Our establishment of model, motivated by actual numerical simulations, can be applied to different situations, with different models.

\subsection{Back to atmospheric background}

This solution also gives an essential theoretical piece of the physical explanation for the behavior of radiation in Biello's system that may lead to a deeper understanding of solitary interaction within the system.

In atmosphere-ocean community, our work can be thought as asymptotic analysis of the nonlinear traveling waves \cite{chen2015nonlinear,ji2014interhemispheric,khouider2012climate} as soliton-like solutions leave behind small scale features after interaction. Furthermore, implications of these structures and interactions for atmospheric tropical/midlatitude behavior is necessary \cite{biello2009nonlinearly}. This sort of nonlinear interaction involving equatorial baroclinic and barotropic Rossby modes might be related to diurnal variations of deep convection in the tropics \cite{raupp2010interaction}.

Back to atmospheric background, our work can be treated as asymptotic analysis of the nonlinear traveling waves \cite{chen2015nonlinear,ji2014interhemispheric,khouider2012climate} as soliton-like solutions leave behind small scale features. Furthermore, much more work is necessary in order to understand the implications of interactions for atmospheric tropical/midlatitude connections \cite{biello2009nonlinearly}. This sort of nonlinear interaction involving equatorial baroclinic and barotropic Rossby modes might be directly related to diurnal variations of deep convection in the tropics \cite{raupp2010interaction}.

\section{Acknowledgments}

Thanks for Trevor Halsted's early numerical experiments showing radiation after interaction. The first author thanks Prof. Biello for two-year patient guidance as well as his parents and Nanjing University (undergraduate institution) for financial and official/ legitimate support on five-month exchange opportunity to UC Davis in 2014.
\setcounter{section}{1}
\appendix
\section{Processing of simulation data to plot figures}
\label{app:processing of simulation data to plot figures}
As for 
\begin{itemize}
\item figure \ref{fig:snapshots of U1 descry}, we plot $u(x,X,T) -K(x)$ due to (\ref{eq:U fleet from simulation});
\item figure \ref{fig:snapshots of U1 scrutine}, $ \epsilon^{-1} \left( U^{F}+\epsilon U^{B}\right)(x,X,T)  = \epsilon^{-1}\left(u(x,X,T )\cosh^2 (x)+12\right) + o(\epsilon^0)$ in simulation with $\epsilon=10^{-2}$.
\item figure \ref{fig:snapshots of G} and part of figure \ref{fig:Schwartz solution G1}, $\frac{U^{F}(x,X,T)}{\sech^2(x)}$  are attained from simulation via \begin{eqnarray*}
&& \frac{U^{F}(x,X,T)}{\sech^2(x)} \\
= &&  \frac{u(x,X,T) -\left(K(x) +o(\epsilon^0)\right)}{\sech^2(x)}
=   \frac{u(x,X,T)+ 12\sech^2(x) }{\sech^2(x)}+o(\epsilon^0) \\
= && \cosh^2 (x)u(x,X,T) +12+o(\epsilon^0).
 \end{eqnarray*}
\item figure \ref{fig:heaviside structure H2}, $\frac{V^{R}(x,X,T)}{R(x)}$ are attained from simulation data by \begin{eqnarray*}
&& \frac{V^{R}(x,X,T)}{R(x)} = \epsilon^{-1} \frac{\epsilon V^{R}(x,X,T)}{R(x)} \\
 = && \frac{\epsilon^{-1}}{R(x)}\left\{ v(x,X,T)-\left[\epsilon^{-2}V^{0}(X,T)+V^{F}(x,X,T)+o(\epsilon^0)\right]\right\}\\
= && \frac{\epsilon^{-1}}{R(x)}\left\{ v(x,X,T)-\left[\epsilon^{-2}V^{0}(X,T)+V^{F}(x,X,T)\right]\right\}+o(\epsilon^0)\\
\simeq &&  \frac{\epsilon^{-1}v(x,X,T)}{R(x)}.
\end{eqnarray*}
Above approximation is admissible just because $\epsilon^{-2}V^{0}(X,T)+V^{F}(x,X,T)$ are narrow enough not to affect our observations;

 \item figure \ref{fig:heaviside structure G2}, $\frac{U^{B}(x,X,T)}{R(x)}$ is attained from simulation via \begin{eqnarray*}
&&\frac{U^{B}(x,X,T)}{R(x)}=  \frac{\epsilon^{-1}}{R(x)}\cdot \epsilon U^{B}(x,X,T) \\
= && \frac{\epsilon^{-1}}{R(x)}\cdot \left\{ u(x,X,T)-\left[K(x)+U^{F}(x,X,T)+o(\epsilon^1)\right]\right\}\\
= &&\frac{\epsilon^{-1}}{R(x)}\left\{ u(x,X,T)-\left[K(x)+U^{F}(x,X,T)\right]\right\}+o(\epsilon^0)\\
 \simeq  && \frac{\epsilon^{-1}\left[ u(x,X,T)-K(x)\right]}{R(x)} +o(\epsilon^0) .
\end{eqnarray*}
Above approximation is admissible since $U^{F}$ is narrow enough not to affect our observations;
\item figure \ref{fig:G2 H2}, part of figure \ref{fig:Schwartz solution G1} and second part of figure \ref{fig:H at xi=0.0 and the one from WKB}, are plotted from data via finite difference method solving Cauchy boundary value problems with two boundary values assigned at $\xi=-10,\xi=10$. 20,000 points are considered.
\end{itemize}
\section{Derivation of WKB equations in section \ref{sec:WKB equation results}}
\label{app:details of section of WKB}
Since 
\begin{eqnarray*}
&& u_t-u_{xxx}+uu_x (\mbox{ use (\ref{eq:ansatz})})\\
= &&  \epsilon^{-3}U^{F}_{T}+\epsilon^{-2}U^{B}_{T}  -\left[ \epsilon^{-3}U^{F}_{XXX}+3\epsilon^{-2}U^{F}_{xXX}+\epsilon^{-2}U^{B}_{XXX}\right]  +o(\epsilon^{-2}) \\
=&& 
 \epsilon^{-3}\cL(U^{F})+\epsilon^{-2}\left[\cL(U^{B})-3U^{F}_{xXX}\right]+ o(\epsilon^{-2}),
\end{eqnarray*}

\begin{eqnarray*}
&& v_t-v_{xxx}+vv_x \\
= &&(\mbox{ use (\ref{eq:ansatz})})\epsilon^{-5}V^{0}_{T}+\epsilon^{-3}V^{F}_{T}+\epsilon^{-2}V^{R}_{T}\\
&&-\left[ \epsilon^{-5}V^{0}_{XXX}+\epsilon^{-3}V^{F}_{XXX}+3\epsilon^{-2}V^{F}_{xXX}+\epsilon^{-2}V^{R}_{XXX}\right] \\
&&+\left[\epsilon^{-2}V^{0}+V^{F}+\epsilon V^{R}\right]\left[\epsilon^{-3}V_{0,X}+\epsilon^{-1}V^{F}_X+(V^{F}_x+V^{R}_X)\right] + o(\epsilon^{-2}) \\
= &&  \epsilon^{-5}V^{0}_{T}+\epsilon^{-3}V^{F}_{T} + \epsilon^{-2}V^{R}_{T}\\
&& -\left[ \epsilon^{-5}V^{0}_{XXX}+\epsilon^{-3}V^{F}_{XXX}  +3\epsilon^{-2}V^{F}_{xXX}+\epsilon^{-2}V^{R}_{XXX}\right] \\
&& +\epsilon^{-5}V^{0}V^{0}_{X}+\epsilon^{-3}(V^{F} V_{0,X}+V^{0}V^{F}_X) \\	
&& +\epsilon^{-2}\left[V^{0}(V^{F}_x+V^{R}_X)+V_{0,X}V^{R}\right]+ o(\epsilon^{-2})
\\
= && 
\epsilon^{-5}\cK(V^{0})+\epsilon^{-3}\left[\cL\left(V^{F}\right)+\left(V^{0}V^{F}\right)_X\right]\\
&& +\epsilon^{-2}\left[\cL(V^{R})-3V^{F}_{xXX}+V^{0}V^{F}_x+\left(V^{0}V^{R}\right)_X\right] + o(\epsilon^{-2}),
\end{eqnarray*}
and
\begin{eqnarray*}
&&(uv)_x\\
= &&(\mbox{ use (\ref{eq:ansatz})}) \left[U^{0}+U^{F}+\epsilon U^{B}\right]\epsilon^{-3}V^{0}_{X}\\
&& +\left[\epsilon^{-1}U^{F}_X  +U^{0}_x+U^{B}_X)\right]  \epsilon^{-2} V^{0}+o(\epsilon^{-2})\\
= &&  
 \epsilon^{-3}\left[U^{0}V^{0}_{X}+(U^{F}V^{0})_X\right] \\
 &&  +\epsilon^{-2}\left[(U^{B}V^{0})_X+(U^{0}_x+U^{F}_x)V^{0}\right]+o(\epsilon^{-2}),
\end{eqnarray*}
where 
\begin{displaymath}
\cL(f)=f_{T}-f_{XXX},\qquad \cK(f)=\cL(f)+ff_X,
\end{displaymath}
Thus, (\ref{eq:WKB U}, \ref{eq:WKB V}) can be obtained.
\section{Derivation of right boundary values for $H^{U}$ and $H^{V}$}
\label{app:Derivation of right boundary values for HU HV}
By taking $\dps \lim_{\xi\rightarrow\infty}$ on both sides of (\ref{eq:ode for G2}), right boundary value of $H^{U}$ appears as:
\begin{eqnarray}
 \lim_{\xi\rightarrow\infty}H^{U}(\xi) && =-\lim_{\xi\rightarrow\infty}\frac{ 8\sqrt{3}/3\int_{-\infty}^{\xi}[12-S^{U}(\eta)]\sech^2(\eta)d\eta}{4+6\sech^2(\xi)} \nonumber \\
 && = \frac{2\sqrt{3}}{3}\int_{\BBR}[S^{U}(\eta)-12]\sech^2(\eta)d\eta \nonumber \\
\mbox{ use (\ref{eq:ode for G})}&&= -\frac{4\sqrt{3}}{9}\int_{\BBR}S^{U}(\eta)d\eta. \label{eq:right boundary value for G2 use integral of G}
\end{eqnarray}

By taking $\dps \lim_{\xi\rightarrow\infty}$ on both sides of (\ref{eq:ode for H2}), right boundary value of $H^{V}$ appears as:
\begin{eqnarray}
\lim_{\xi\rightarrow\infty}H^{V}(\xi) && = \lim_{\xi\rightarrow\infty}\frac{-8\sqrt{3}/3\int_{-\infty }^{\xi}\left[12-S^{U}(\eta)+2S^{V}(\eta)\right]\sech^2(\eta)d\eta }{4-12\sech^2(\xi)} \cdot \nonumber \\
&& =\frac{2\sqrt{3}}{3}\int_{\BBR}\left[S^{U}(\eta)-12-2S^{V}(\eta)\right]\sech^2(\eta)d\eta \nonumber \\
\mbox{use (\ref{eq:ode for H})}&&= -\frac{4\sqrt{3}}9\int_{\BBR}S^{V}(\eta)d\eta.  \label{eq:right boundary value for H2 use integral of H}
\end{eqnarray}

\section{Observation and verification of $V^{F}$}
\label{app:observation of H}
\begin{figure}[tb]
\includegraphics[width=0.5\textwidth]{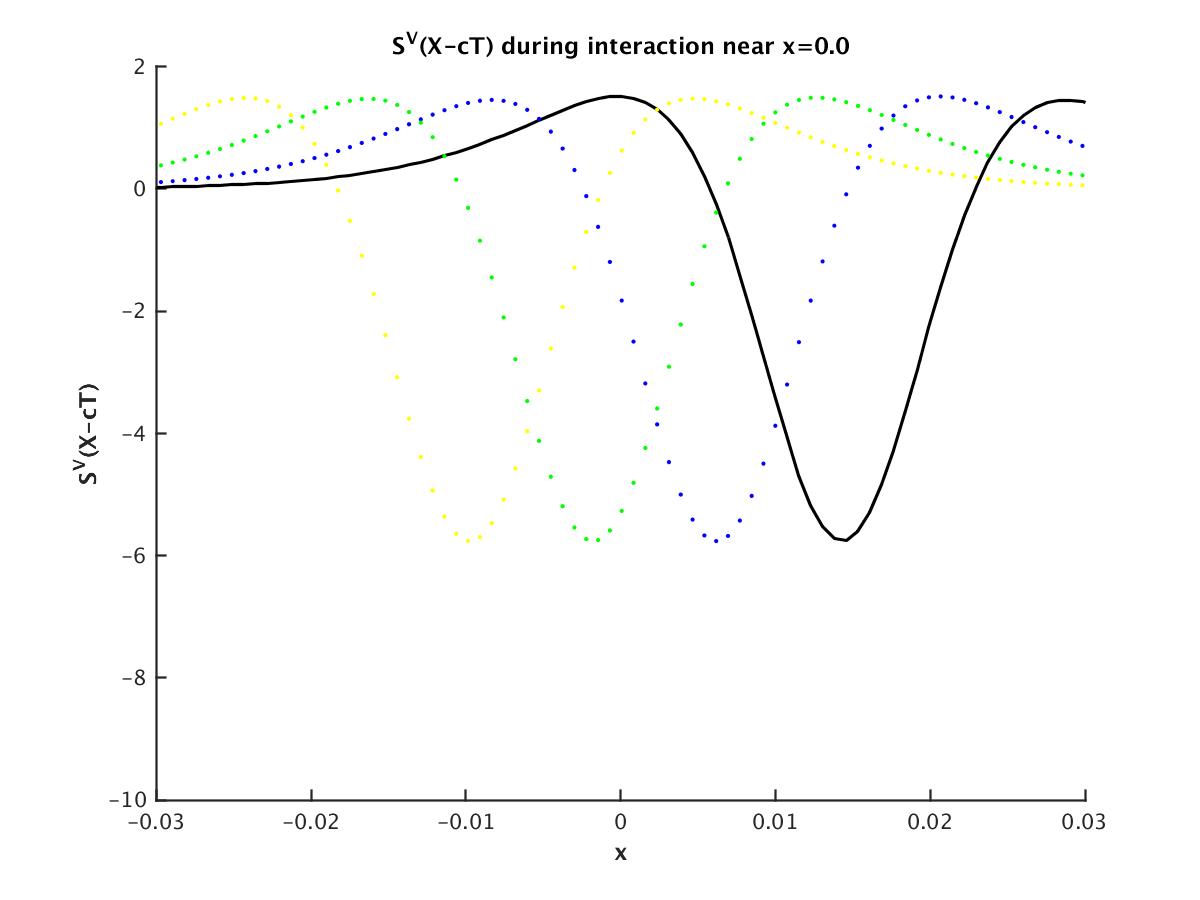}%
\includegraphics[width=0.5\textwidth]{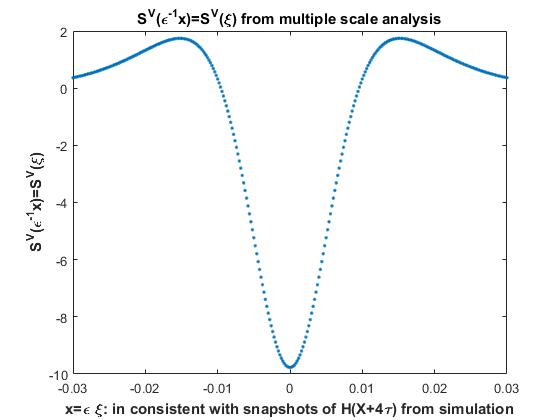}
\caption{First: $S^{V}(\xi)=S^{V}(X-cT)$ at $x\simeq 0$ from simulation with $\epsilon =10^{-2}$; second: $S^{V}\left(\epsilon^{-1}x\right) =S^{V}(\xi)$ from multiple scale analysis with $x=\epsilon \xi$ in order to be consistent with $S^{V}(X-cT)$ from simulation -- it is predicted by asymptotic model from (\ref{eq:ode for H}) and numerically plotted via finite difference method with Cauchy boundary condition $S^{V}(-10)=S^{V}(10)=0$.}
\label{fig:H at xi=0.0 and the one from WKB}
\end{figure}

Numerical observation of $V^{F}$ is quite difficult once realizing $\epsilon^{-2}V^{0}$ part of wave $v$ is so tall  that one could hardly observe its bottom (namely, the area where $v$ obtains $\min v$); this corresponds to the fact that in the limit of $\epsilon \to 0$, $v$ wave tends to be singular. Luckily, we can balance the order of $\epsilon$ in (\ref{eq:nonlinearly coupled KdV equation of Biello}) instead. This reveals $V^{F}$ is of order $O(\epsilon^0)$. 

Again,we can validate our ansatz (\ref{eq:ansatz}) 
$$
v(x,t)=\epsilon^{-2}V^{0}(X,T)+V^{F}(x,X,T)+\epsilon V^{R}(x,X,T)+o(\epsilon^0),
$$
and corresponding fast-slow decomposition (\ref{eq:structure of V1})
\begin{displaymath}
V^{F}(x,X,T)=\sech^2(x)S^{V}(X-cT).\end{displaymath}

Notice from fast-slow decomposition (\ref{eq:structure of V1}) and ansatz (\ref{eq:ansatz}), we have
\begin{eqnarray*}
&& \cosh^2(x) V^{F}(x,X,T) \\
= && \cosh^2(x)\left\{v(x,X,T)-\left[\epsilon^{-2} V^{0}\left(X-\delta(x,X,T),T\right)+o(\epsilon^0)\right]\right\} \\
= && \cosh^2(x)\left\{v(x,X,T)-\epsilon^{-2} F\left(X-\delta(x,X,T),T\right)\right\}+o(\epsilon^0) \\
\end{eqnarray*}
where $\delta(x,X,T)(>0)$ increases as $T(>0)$ increases and $x$ fixed, since existence of $u$ indeed "delays" KdV movement of $v$; by comparison, existence of $v$ indeed "delays" KdV movement of $u$; modification of shifts can only be manually made up till now.

Figure \ref{fig:H at xi=0.0 and the one from WKB} shows 
\begin{itemize}
\item (first) $S^{V}$ from numerical simulations near $x=0$ via the numerical method described above and 
\item (second) $S^{V}$ predicted by asymptotic model from (\ref{eq:ode for H}) numerically via finite difference method with Cauchy boundary condition $S^{V}(-10)=S^{V}(10)=0$.
\end{itemize} 
The solitary structure with two small bumps at two sides is evident and is a shared feature for theoretical results and numerical results. However, the integral quantity as well as other feature for exact size of $S^{V}$ is different. According to (\ref{eq:right boundary value for H2 use integral of H}), it is precisely this integral quantity that determines right boundary value for $H^{V}$, which is an important feature for $H^{V}$. According to (\ref{eq:right boundary value for H2 use integral of H}), the inconsistency can be view of ill-prediction of integral quantity of $S^{V}$ by our asymptotic model.

\section*{References}
\bibliographystyle{iopart-num}
\bibliography{iopscattering}

\end{document}